\documentclass[11pt,a4paper]{article}
\usepackage{jcappub}

\usepackage[utf8]{inputenc}

\hypersetup{allcolors=[rgb]{1,0.56,0}}

\newcommand{\orcid}[1]{ORCID: \href{http://orcid.org/#1}{#1}}
\newcommand{\Edep}{E_{\rm dep}}
\newcommand{\fgal}{f_{\rm gal}}
\newcommand{\hfgal}{\hat f_{\rm gal}}
\renewcommand{\L}{\mathcal L}

\newcommand{\e}[1]{\times10^{#1}}

\newcommand{\ANA}{\texttt{ANA}}

\title{The Galactic Contribution to IceCube's Astrophysical Neutrino Flux}

\author[a,1]{Peter B.~Denton,\note{\orcid{0000-0002-5209-872X}}}
\emailAdd{peterbd1@gmail.com}
\author[b]{Danny Marfatia,}
\emailAdd{dmarf8@hawaii.edu}
\author[c]{Thomas J.~Weiler}
\emailAdd{tom.weiler@vanderbilt.edu}

\affiliation[a]{Niels Bohr International Academy, University of Copenhagen, The Niels Bohr Institute, Blegdamsvej 17, DK-2100, Copenhagen, Denmark}
\affiliation[b]{Department of Physics and Astronomy, University of Hawaii at Manoa, Honolulu, HI 96822, USA}
\affiliation[c]{Department of Physics \& Astronomy, Vanderbilt University, Nashville, TN 37235, USA}

\abstract{High energy neutrinos have been detected by IceCube, but their origin remains a mystery.
Determining the sources of this flux is a crucial first step towards multi-messenger studies.
In this work we systematically compare two classes of sources with the data: Galactic and extragalactic.
We assume that the neutrino sources are distributed according to a class of Galactic models.
We build a likelihood function on an event by event basis including energy, event topology, absorption, and direction information.
We present the probability that each high energy event with deposited energy $E_{\rm dep}>60$ TeV in the HESE sample is Galactic, extragalactic, or background.
For Galactic models considered the Galactic fraction of the astrophysical flux has a best fit value of $1.3\%$ and is $<9.5\%$ at 90\% CL.
A zero Galactic flux is allowed at $<1\sigma$.}


\arxivnumber{1703.09721}

\begin{document}

\maketitle

\section{Introduction}
\label{sec:Introduction}
IceCube has reported the detection of high energy astrophysical neutrinos for the first time \cite{Aartsen:2013bka}.
This opens up a new channel to probe the nature of the extreme universe that cannot be directly reproduced in the laboratory \cite{Anchordoqui:2013dnh}.
By combining information from neutrinos with information from electromagnetic radiation, cosmic rays, and gravitational waves, we can begin to form a detailed picture of the microscopic properties of the most extreme objects and environments in nature.

IceCube has now released information on 82 neutrinos with energies $\gtrsim20$ TeV whose initial interaction points are contained inside the detector.
These events comprise the four year high-energy starting event (HESE) catalog \cite{Aartsen:2014gkd,Aartsen:2015zva,IC:ICRC17}.
While it is known that many of the neutrinos measured are of astrophysical origin, the nature of the sources is unknown despite numerous searches.
The cleanest method to determine the sources of astrophysical neutrinos is via a point source search using multiple track events with good angular resolution coming from the same direction.
Thus far no point sources have been found by IceCube or ANTARES \cite{Abbasi:2009cv,Adrian-Martinez:2014wzf,Aartsen:2014cva,Adrian-Martinez:2015ver,Aartsen:2016tpb}, although prospects may improve if the high energy extension IceCube-Gen2 is built \cite{Mertsch:2016hcd}.

Motivated by a cascade event with energy $E_\nu\sim1$ PeV and central direction $1.2^\circ$ from the Galactic center \cite{Bykov:2015nta,Celli:2016uon} and a median angular uncertainty of $13.2^\circ$, there has been significant interest in determining if there is a Galactic component to the astrophysical neutrino flux.
The first natural thing to test is the number of events within a certain window of the Galactic plane \cite{Stecker:1978ah,Aartsen:2013jdh,Ahlers:2013xia,Aartsen:2014gkd,Anchordoqui:2014rca,Murase:2015xka,Neronov:2015osa,Troitsky:2015cnk,Chianese:2016opp,Padovani:2016wwn,Palladino:2016zoe,Palladino:2016xsy,IC:ICRC17}.
This has shown a weak suggestion that there may be an excess of events within a window of $7.5^\circ$ of the Galactic plane at $p=0.028$.
When the appropriate trial factor is included for scanning over opening angles in galactic coordinates, the significance is reduced to $p=0.24$.
ANTARES has also performed a scan for anisotropies finding an excess in the direction of the Galactic center with significance $2.1\sigma$ after applying multiple trial factors including scanning over the sky \cite{Albert:2017fvi}.

An alternative method to scans is to use information about the shape of the Galaxy and information about specific Galactic sources to avoid penalty factors.
Some approaches have been to consider contributions from Galactic cosmic rays, constraints from gamma rays measured by Fermi, neutrinos from the Galactic center, various Galactic catalogs such as pulsars, pulsar wind nebulae, supernova remnants, decay/absorption from the dark matter halo, the Cygnus-X region, bright nearby stars, and even exoplanets \cite{Aartsen:2014cva,Ahlers:2015moa,Bykov:2015nta,Emig:2015dma,Troitsky:2015cnk,Kistler:2015oae,Fang:2015xhg,Sahakyan:2015bgg,Padovani:2016wwn,Marinelli:2016mjo,Celli:2016uon,Arguelles:2017atb,Yoast-Hull:2017gaj}.
In all cases it was found that a single Galactic component cannot explain the entirety of the astrophysical flux.

Numerous classes of extragalactic sources have also been considered to explain the IceCube flux.
Active galactic nuclei (AGNs), blazars, star forming galaxies (SFGs), and gamma ray bursts (GRBs) have been often considered as the sources by both IceCube and others \cite{Stecker:1991vm,Stecker:2005hn,Loeb:2006tw,Anchordoqui:2014yva,Aartsen:2014cva,Zandanel:2014pva,Murase:2015xka,Bechtol:2015uqb,Kistler:2015ywn,Murase:2015ndr,Aartsen:2016qcr,Padovani:2016wwn,Moharana:2016yoy,Murase:2016gly,Feyereisen:2016fzb,Neronov:2016ksj,Aartsen:2017wea}.
These models often have difficulties in fitting the entire HESE data.
Some models are only able to fit a portion of the spectrum, while others find a difficulty with simultaneously fitting Fermi high energy gamma ray data and IceCube data for a given source model.
In addition, correlations with UHECRs, which are generally believed to be extragalactic, shows no significant correlation with HESE neutrinos \cite{Aartsen:2015dml}.

Alternatively there have been some claims of correlations of the HESE data with specific sources, the majority of which are with blazars, although typically of weak statistical significance \cite{Kadler:2016ygj,Halzen:2016uaj,Kun:2016bnk,Gao:2016uld}.
Even if these are sources of certain events, they may still not be the source of the diffuse neutrino flux.

We take an intermediate approach in this paper.
We use information about the shape of the Galaxy, but avoid considering multiple Galactic models that are not statistically independent, by considering sources distributed throughout the Galaxy that follow the matter distribution in the Galaxy.
Then, within the angular resolution of IceCube, a majority of Galactic distributions should be consistent with this template.\footnote{\label{fn:MW Shape}Three exceptions of interest are the Fermi bubbles, the Crab Nebula, and an extended dark matter halo.
The constraints on Galactic emission from Fermi bubbles from ref.~\cite{Ahlers:2015moa} are the strongest of all the Galactic models considered in that paper at $<25\%$ of the flux at $90\%$ CL.
Moreover, while there appears to be a slight excess in the southern Fermi bubble (some of which could be from sources in the Galactic plane due to the large angular uncertainty of the cascade events), the northern Fermi bubble appears to be in a deficit, likely contributing to the stronger constraints.
Finally, HAWC measurements disfavor hadronic models which would produce high energy neutrinos in the Fermi bubbles \cite{Abeysekara:2017wzt}.
The constraints on the high energy neutrino flux from the Crab Nebula are very strong as they were taken during a flare and an under fluctuation was recorded \cite{IceCube:2011aa}.
Dark matter and its mode for producing high energy astrophysical neutrinos may not exist, let alone contribute to the event sample we use.}
The outline of the paper is as follows.
In section \ref{sec:Data} we present the four year HESE dataset reported by IceCube.
Section \ref{sec:Anisotropy} describes how we quantitatively compare different distributions with the data.
We then create a likelihood by combining information about topology, absorption, energy, and direction in section \ref{sec:Likelihood}.
The results are then presented in \ref{sec:Results} followed by conclusions in \ref{sec:Discussion}.
The results of this paper were generated with the Astrophysical Neutrino Anisotropy package \ANA\footnote{Available at \url{https://github.com/PeterDenton/ANA}.} \cite{peter_b_denton_2017_438675}.

\section{IceCube HESE Dataset}
\label{sec:Data}
The six year HESE dataset contains 82 events with deposited energies $>20$ TeV up to $\sim$2 PeV with zero events due to the Glashow resonance at $E_{\bar\nu_e}\sim6.3$ PeV \cite{Glashow:1960zz}.
The deposited energy corresponds to the true neutrino energy only for cascades from CC interactions of $\nu_e$ and $\nu_\tau$.
CC $\nu_\mu$ interactions lead to track topologies, for which a correction factor needs to be applied to estimate the true neutrino energy \cite{Anchordoqui:2016ewn}.
We impose a higher cutoff of $\Edep>60$ TeV to remove the majority of the backgrounds from muons and atmospheric neutrinos, both of which come from extensive air showers (EAS's) created by cosmic ray interactions in the atmosphere.
This cut improves the purity of the sample.

\begin{figure}
\centering
\includegraphics[width=\textwidth]{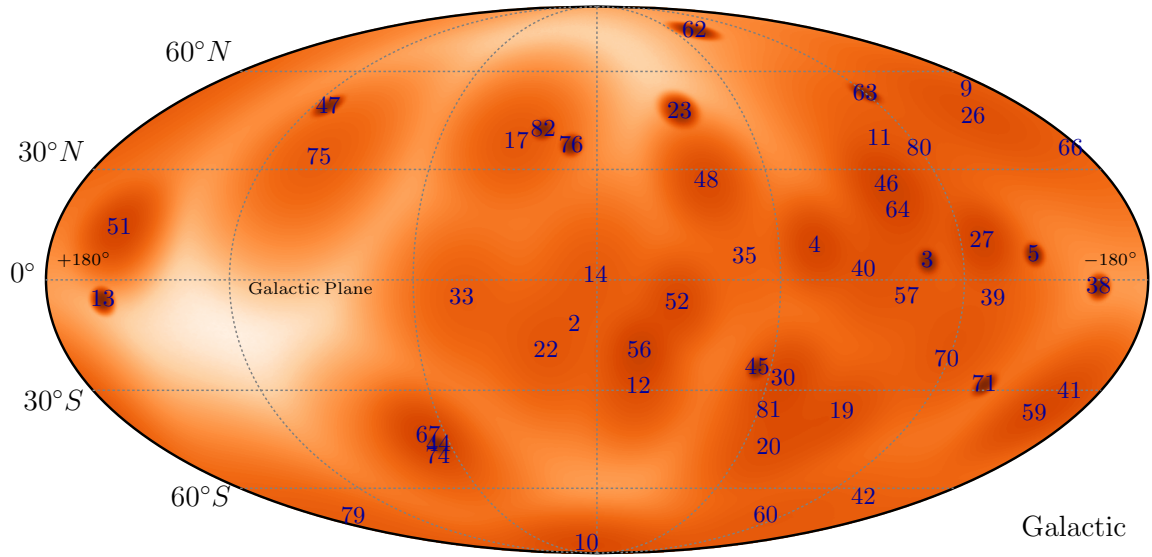}
\caption{The locations of the 50 events with $E_{\rm dep}>60$ TeV and the distribution of likely arrival directions accounting for the angular uncertainty using the von Mises-Fisher distribution.
The events are plotted in galactic coordinates with the Galactic center in the center of the figure and longitude increasing to the left in a Mollweide projection.
The apparent deficit of events in the left hand region of the figure is due to the fact that at these energies the Earth absorbs a significant amount of the up-going neutrino flux.}
\label{fig:SkyMap}
\end{figure}

Several specific events in the catalog are of particular note.
Events 32 and 55 are both coincident events, with what appears to be two separate interactions overlapping in time, and are almost certainly the result of EAS's.
We do not include them in the sample because no energy can be reconstructed from such an event.
Event 28 is also likely a background event, probably an atmospheric muon, although with deposited energy $\Edep=46.1$ TeV below our cut.
Both events 28 and 32 triggered IceTop, the cosmic ray detector located on the surface of the ice, at a subthreshold level \cite{Aartsen:2014gkd}.
Finally, event 45 is a down going track event at declination $-86.3^\circ$ and, due to a lack of hits on IceTop, is very likely not the result of an EAS; we take it to be astrophysical.
In total, we have 50 events in our final sample shown in fig.~\ref{fig:SkyMap} including an event that is certainly astrophysical (\#45).

We note that the issue of event topology misidentification (tracks misidentified as showers) for background events would have a negligible effect on our results since the numbers of both kinds of events are quite small for $\Edep>60$ TeV \cite{Aartsen:2014gkd,Palomares-Ruiz:2015mka}.

Another high energy data set reported by IceCube is the through going track data set \cite{Aartsen:2016xlq} containing high energy tracks, likely from $\nu_\mu$'s, that start outside the detector.
To reduce the massive atmospheric backgrounds, IceCube only considers up going (Northern hemisphere) tracks.
IceCube has reported a tension in simultaneously fitting the spectrum to both the HESE data and the through going track data.
Some use this tension of the track fit with the pure HESE events to claim $3\sigma$ evidence for a break in the single power law fit \cite{Anchordoqui:2016ewn}.
Our study is immune to this controversy as we will focus only on the HESE data set; this focus prevents systematic uncertainties from dwarfing the statistics.

\section{Galactic and Extragalactic Distributions}
\label{sec:Anisotropy}
\begin{figure}
\centering
\includegraphics[width=\textwidth]{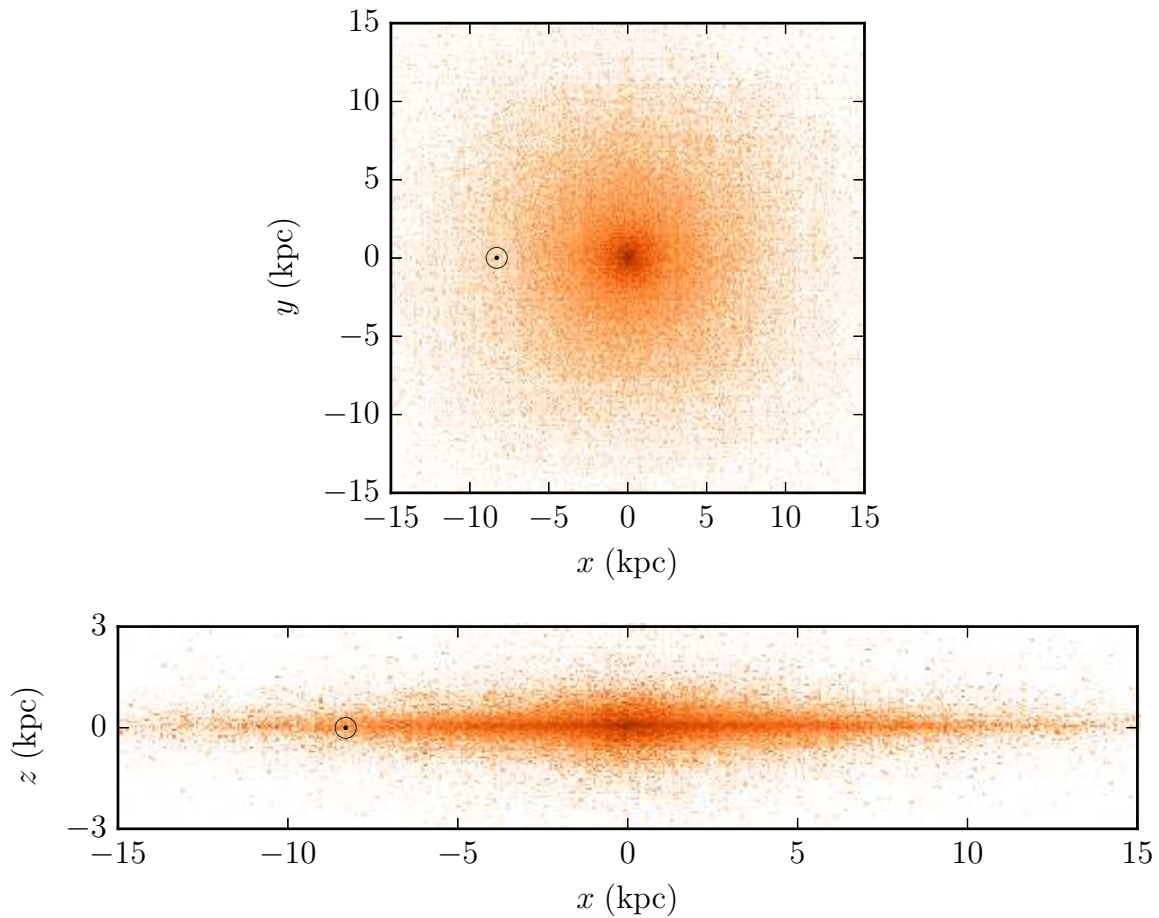}
\caption{The top down (above) and edge on (below) view of the matter distribution of the Galaxy $\rho_{\rm gal}(\mathbf{x})$ described in \cite{McMillan:2011wd}.
The location of our solar system is denoted with $\odot$.}
\label{fig:MW Visualization}
\end{figure}

\begin{figure}
\centering
\includegraphics[width=\textwidth]{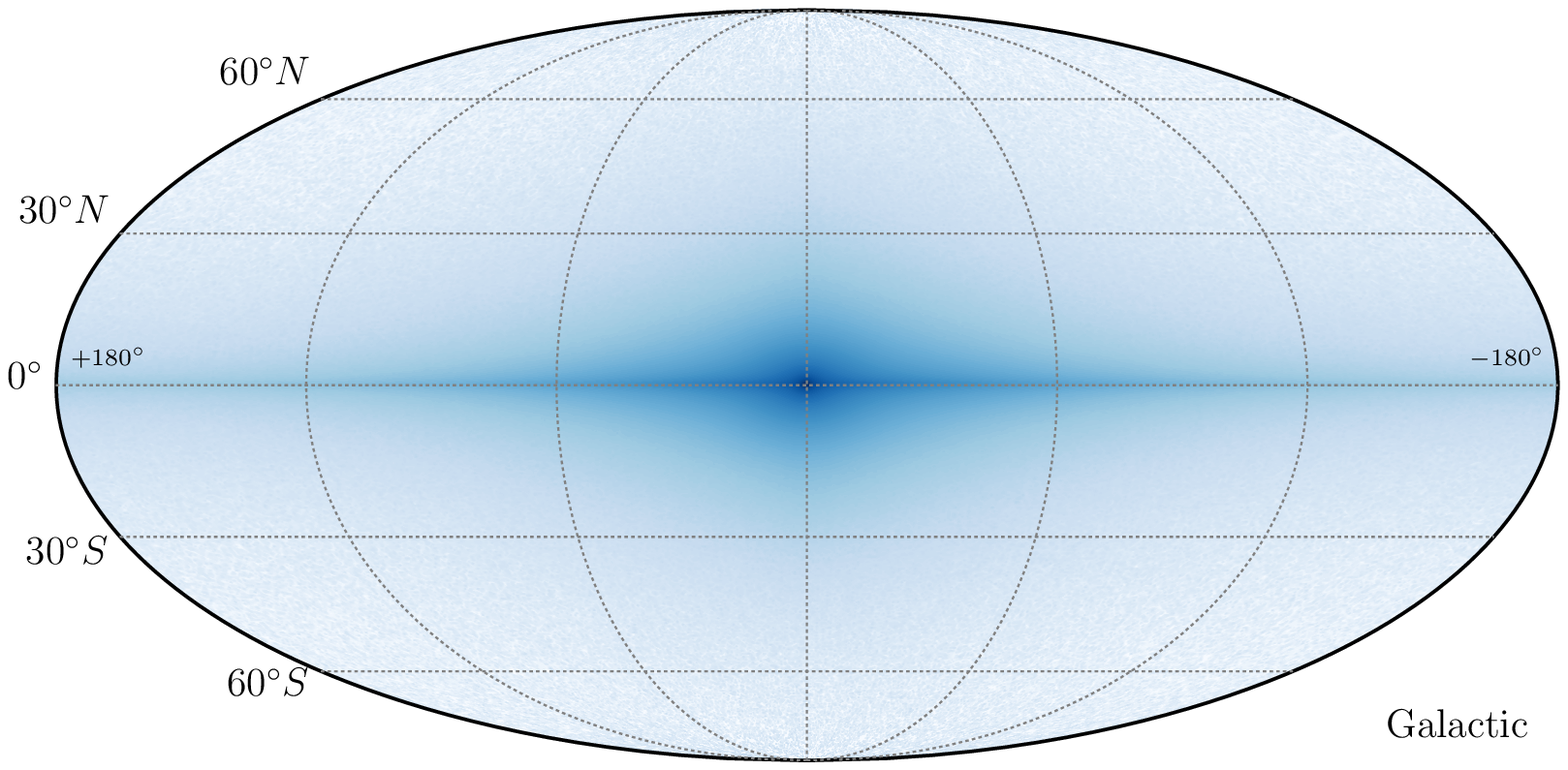}
\caption{The sky map of the Galactic distribution in eq.~\ref{eq:Phigal} with parameters described in \cite{McMillan:2011wd} shown in galactic coordinates.
The model consists of a central axisymmetric bulge, a thin disc, and a thick disc.}
\label{fig:MW SkyMap}
\end{figure}

We consider astrophysical sources distributed according to one of two distributions: Galactic and extragalactic.
Since extragalactic sources are generally expected to contribute to a diffuse flux of neutrinos, we treat the extragalactic component as purely isotropic with a probability density function (pdf) $\Phi_{\rm exgal}(\Omega)=\frac1{4\pi}$.
For the Galactic flux we consider sources distributed according to the matter in the Galactic plane as parameterized in \cite{McMillan:2011wd}.
This model contains an axisymmetric bulge in the center of the Galaxy, a thin disc, and a thick disc.
We use the best fit values which leads to the distribution of sources shown in fig.~\ref{fig:MW Visualization} and the skymap shown in fig.~\ref{fig:MW SkyMap}.
We note that, as expected, our results are independent of the details of the Galactic parameterization.
Then the Galactic pdf\footnote{This is analogous to the $D$-factors sometimes used in dark matter decay analyses.} is,
\begin{equation}
\Phi_{\rm gal}(\Omega)=\frac{\int ds\,\rho_{\rm gal}(s,\Omega)}{\int dsd\Omega'\,\rho_{\rm gal}(s,\Omega')}\,,
\label{eq:Phigal}
\end{equation}
where $s$ is the line of sight and $\Omega$ is the angular direction, both taken in the reference frame centered on our sun.
Our reference frame and the Galactic reference frame are related by a shift of $8.29$ kpc from the Galactic center.

Since there is Milky Way in every direction from Earth, i.e.~$\Phi_{\rm gal}(\Omega)$ is nonzero $\forall\,\Omega$, every event has some chance of being Galactic.
Every event also has some chance of being extragalactic.
Far away from the Galactic plane, where $\Phi_{\rm gal}<\frac1{4\pi}$, events are more likely to be extragalactic than Galactic, and the presence of events there will push the likelihood (see section \ref{sec:Likelihood}) to prefer a greater extragalactic component, and vice versa for events near the galactic plane and the galactic center.
Thus our results may depend somewhat on our definition of the Milky Way Galaxy (see e.g.~the caveats in footnote \ref{fn:MW Shape}).

\section{Likelihood}
\label{sec:Likelihood}
To determine how well our model of two components fit the data, we construct a likelihood function with $\fgal$ as a free parameter, where $\fgal$ is the fraction of the astrophysical flux, not including backgrounds, from the Galaxy, as seen at IceCube.

The likelihood that an event is a background is taken from \cite{Aartsen:2014gkd} as described in \cite{Ahlers:2015moa}.
We use $\Edep>60$ TeV, at which point we expect $0.85$ muon events as contamination and $5.04$ neutrinos from atmospheric interactions.
We then take the distribution of backgrounds as a function of the deposited energy, declination, and topology.
The expected number of background events due to muons and neutrinos in our sample is,
\begin{equation}
N_{\rm bkg}(\Edep,\delta,t)=x_{\mu,t}\phi_\mu(\Edep,\delta)+x_{\rm atm,t}\phi_{\rm atm}(\Edep,\delta)\,,
\end{equation}
where $x_{\mu,C}=0$, $x_{{\rm atm},C}=1$, $x_{\mu,T}=0.19$, $x_{{\rm atm},T}=0.81$ for cascade and track topologies.
The $\phi(\Edep,\delta)$ functions are the expected number of $\mu$ or atmospheric events in a given $\Edep$, $\delta$ bin taken from \cite{Aartsen:2014gkd}.
$N_{\rm astro}=\phi_{\rm astro}(\delta)$ is similarly defined using the declination information from the same reference.
Then the likelihood that event $i$ is a background or astrophysical is,
\begin{align}
\L_{{\rm bkg},i}&=N_{\rm bkg}(E_{{\rm dep},i},\delta_i,t_i)\,,\\
\L_{{\rm astro},i}&=N_{\rm tot}(E_{{\rm dep},i},\delta_i,t_i)-N_{\rm bkg}(E_{{\rm dep},i},\delta_i,t_i)\,,
\end{align}
where $N_{\rm tot}(E_i,\delta_i,t_i)$ is the total number of measured events in the same deposited energy bin as event $i$, the same declination bin as event $i$, and with the same topology as event $i$, and the energy uncertainty ($\sim$8--$14\%$) is marginalized over.
These likelihoods are defined up to an overall normalization factor that is irrelevant since we will be using a log likelihood ratio as our test statistic.

In order to determine if an astrophysical neutrino is likely to be Galactic or extragalactic, our protocol is to calculate the conditional likelihoods based on direction information and the Galactic matter distribution.
The likelihoods are then given by,
\begin{align}
\L_{{\rm gal}|{\rm astro},i}(\fgal)&=\fgal\int d\Omega\,\Phi_{\rm gal}(\Omega)f_{\rm vMF}(\theta,\kappa_i)\,,\\
\L_{{\rm exgal}|{\rm astro},i}(\fgal)&=(1-\fgal)\frac1{4\pi}\,.
\label{eq:conditional likelihoods}
\end{align}
$f_{\rm vMF}$ is the von Mises-Fisher distribution, and $\kappa_i$ is the concentration of event $i$ which is related to the reported median angular uncertainty $\alpha_{50\%}$ as described in appendix \ref{sec:vMF}.
$\theta$ is the angle between $\Omega$ and the best fit direction of event $i$.
Since the astrophysical component is normalized to an integral over opening angles, we similarly normalize the background component with an additional factor of $\frac1{4\pi}$ \cite{Sergio:2017}.

Then the total likelihood is,
\begin{equation}
\L(\fgal)=\prod_i\L_i(\fgal)\,,
\label{eq:total likelihood}
\end{equation}
where,
\begin{equation}
\L_i(f_{\rm gal})=\left[\L_{{\rm gal}|{\rm astro},i}(\fgal)+\L_{{\rm exgal}|{\rm astro},i}(\fgal)\right]\L_{{\rm astro},i}+\frac1{4\pi}\L_{{\rm bkg},i}\,.
\label{eq:likelihood one event}
\end{equation}
We then calculate this likelihood for $\fgal\in[0,1]$ to find $\hfgal$ which maximizes the likelihood.

\section{Results}
\label{sec:Results}
\begin{figure}
\centering
\includegraphics[width=3.5in]{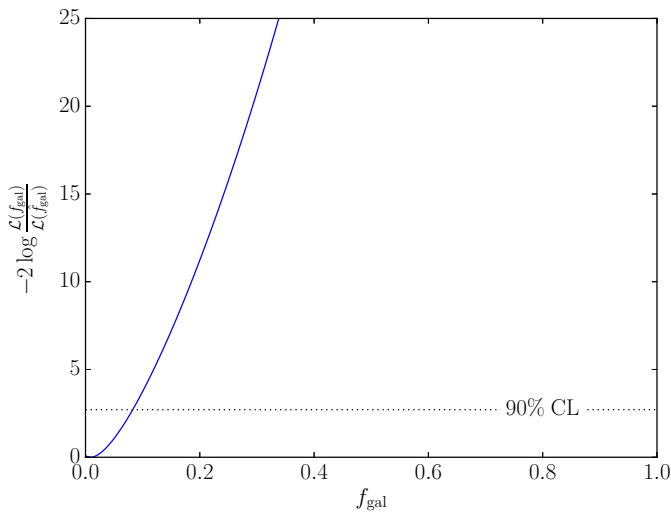}
\caption{The log likelihood ratio scan $-2\log\mathcal L(f_{\rm gal})/\mathcal L(\hat f_{\rm gal})$.
We find the best fit point at $\hat f_{\rm gal}=0.013$ and $f_{\rm gal}=0$ is allowed at $<1\sigma$.}
\label{fig:Likelihood}
\end{figure}

We use the total likelihood function described in eq.~\ref{eq:total likelihood}.
We sample the Galactic distribution using \ANA~\cite{peter_b_denton_2017_438675} that calculates the likelihoods.
A scan of likelihoods over $\fgal$ is shown in fig.~\ref{fig:Likelihood}.
We plotted the test statistic $TS=-2\log\L(\fgal)/\L(\hfgal)$ which can be reasonably estimated to follow a $\chi^2$ distribution with one degree of freedom by Wilks' theorem \cite{wilks1938}.
We find $\fgal=0.013$ with exclusion ranges and upper limits on the Galactic contribution to the astrophysical flux shown in table \ref{tab:limits}.
$\fgal=0$ is allowed at $0.44\sigma$ and $\fgal=1$ is excluded at $16\sigma$.
At $>5\sigma$ significance, the event sample is dominated by extragalactic sources.

\begin{table}
\centering
\caption{Confidence intervals and upper limits for the Galactic contribution to the astrophysical flux at various common confidence levels.
The best fit value is $\hat f_{\rm gal}=0.013$ and $f_{\rm gal}=0$ is slightly allowed at $<1\sigma$.}
\label{tab:limits}
\begin{tabular}{c|c}
CL			&$\fgal$\\\hline
$1\sigma$ & $<0.057$\\
$90\%$ & $<0.095$\\
$2\sigma$ & $<0.12$\\
$3\sigma$ & $<0.2$\\
$4\sigma$ & $<0.28$\\
$5\sigma$ & $<0.38$
\end{tabular}
\end{table}

The probability that event $i$ is of Galactic origin, of extragalactic origin, or is atmospheric background, is calculated by,
\begin{align}
p_{{\rm gal},i}&=\frac{\L_{{\rm gal}|{\rm astro},i}(\hfgal)\L_{{\rm astro},i}}{\L_i(\hfgal)}\,,\\
p_{{\rm exgal},i}&=\frac{\L_{{\rm exgal}|{\rm astro},i}(\hfgal)\L_{{\rm astro},i}}{\L_i(\hfgal)}\,,\\
p_{{\rm bkg},i}&=\frac{\frac1{4\pi}\L_{{\rm bkg},i}}{\L_i(\hfgal)}\,.
\end{align}
Table \ref{tab:event probabilities} lists each probability for each event, given our Galactic definition.
As a check, we note that the probabilities for events to have a Galactic origin increase for events pointing toward the Galactic plane, and especially increase for those events pointing toward the Galactic center.
The best fit number of Galactic, extragalactic, and background events within the sample is then calculated by summing the probabilities.
We get for our sample of 50 events,
\begin{equation}
\sum_ip_{{\rm gal},i}=0.6\,,\qquad\sum_ip_{{\rm exgal},i}=45.3\,,\qquad\sum_ip_{{\rm bkg},i}=4.1\,.
\end{equation}
The events that are most likely to be Galactic with $p_{\rm gal}>0.01$ are 2, 14, 22, 33, and 52.

\begin{table}
\centering
\caption{The probability that each event is Galactic, extragalactic, and a background, evaluated for $\hat f_{\rm gal}=0.013$.
Events are ordered according to their true energies in TeV.
Values less than $1\e{-5}$ are set to zero.}
\label{tab:event probabilities}
\begin{tabular}{c|c|c|c|c}
$E$&id&$p_{\rm gal}$&$p_{\rm exgal}$&$p_{\rm bkg}$\\\hline
$2003$ & 35 & $0.0096$ & $0.99$ & $0$\\
$1140$ & 20 & $2\e{-5}$ & $1$ & $0$\\
$1040$ & 14 & $0.36$ & $0.64$ & $0$\\
$885$ & 45 & $1.2\e{-4}$ & $1$ & $0$\\
$512$ & 13 & $1.8\e{-4}$ & $1$ & $8.6\e{-4}$\\
$404$ & 38 & $3.8\e{-4}$ & $0.87$ & $0.13$\\
$384$ & 33 & $0.012$ & $0.98$ & $0.0045$\\
$318$ & 82 & $2.7\e{-5}$ & $0.56$ & $0.44$\\
$249$ & 76 & $6.8\e{-5}$ & $0.7$ & $0.3$\\
$219$ & 22 & $0.046$ & $0.93$ & $0.021$\\
$210$ & 26 & $0$ & $0.88$ & $0.12$\\
$199$ & 17 & $1.9\e{-4}$ & $0.84$ & $0.16$\\
$190$ & 63 & $1.1\e{-5}$ & $0.75$ & $0.25$\\
$165$ & 67 & $0$ & $0.47$ & $0.53$\\
$165$ & 4 & $0.0017$ & $1$ & $0$\\
$164$ & 44 & $1.4\e{-5}$ & $0.84$ & $0.16$\\
$164$ & 75 & $4.2\e{-5}$ & $1$ & $0$\\
$159$ & 23 & $2.8\e{-5}$ & $0.94$ & $0.06$\\
$158$ & 79 & $0$ & $0.81$ & $0.19$\\
$158$ & 52 & $0.043$ & $0.96$ & $0$\\
$158$ & 46 & $4.2\e{-5}$ & $0.94$ & $0.057$\\
$157$ & 40 & $0.0014$ & $1$ & $0$\\
$152$ & 3 & $4.7\e{-4}$ & $0.95$ & $0.046$\\
$151$ & 81 & $1.2\e{-4}$ & $1$ & $0$\\
$146$ & 62 & $0$ & $0.89$ & $0.11$
\end{tabular}\,
\begin{tabular}{c|c|c|c|c}
$E$&id&$p_{\rm gal}$&$p_{\rm exgal}$&$p_{\rm bkg}$\\\hline
$143$ & 47 & $0$ & $0.96$ & $0.041$\\
$141$ & 71 & $1.6\e{-5}$ & $0.92$ & $0.079$\\
$137$ & 5 & $1.3\e{-4}$ & $0.81$ & $0.19$\\
$132$ & 57 & $6.9\e{-4}$ & $1$ & $0$\\
$128$ & 30 & $1\e{-4}$ & $1$ & $0$\\
$124$ & 59 & $0$ & $0.81$ & $0.19$\\
$117$ & 2 & $0.12$ & $0.87$ & $9.5\e{-4}$\\
$104$ & 48 & $3.2\e{-4}$ & $1$ & $0.0032$\\
$104$ & 56 & $0.0046$ & $1$ & $0$\\
$104$ & 12 & $0.002$ & $1$ & $0$\\
$101$ & 39 & $2.8\e{-4}$ & $0.96$ & $0.04$\\
$98$ & 70 & $9.9\e{-5}$ & $0.99$ & $0.0064$\\
$97$ & 10 & $0$ & $0.99$ & $0.0074$\\
$93$ & 60 & $0$ & $1$ & $0$\\
$88$ & 11 & $3.9\e{-5}$ & $0.9$ & $0.095$\\
$87$ & 41 & $1.4\e{-5}$ & $0.78$ & $0.22$\\
$85$ & 80 & $3.5\e{-5}$ & $0.91$ & $0.091$\\
$84$ & 66 & $2.5\e{-5}$ & $0.95$ & $0.054$\\
$76$ & 42 & $0$ & $0.98$ & $0.017$\\
$71$ & 19 & $2.6\e{-5}$ & $1$ & $0$\\
$71$ & 74 & $1.6\e{-5}$ & $0.77$ & $0.23$\\
$70$ & 64 & $1.9\e{-4}$ & $0.98$ & $0.016$\\
$66$ & 51 & $6.3\e{-5}$ & $0.96$ & $0.044$\\
$63$ & 9 & $0$ & $0.91$ & $0.092$\\
$60$ & 27 & $1.8\e{-4}$ & $0.89$ & $0.11$
\end{tabular}
\end{table}

\section{Discussion and Conclusions}
\label{sec:Discussion}
We have estimated the origin of IceCube's astrophysical neutrino flux by considering sources split into two categories.
An extragalactic flux of neutrinos is likely dominated by a diffuse flux, as gleaned from the null result of point source searches
(to date, auto-correlation studies as well as cross-correlation studies have failed to reveal any point sources of neutrinos).
We model the extragalactic flux, accordingly, as isotropic.
We model the Galactic flux by the known distribution of matter in the Galaxy \cite{McMillan:2011wd}.
The Galactic flux will likely follow a distribution that is well approximated by the distribution of matter in the Galaxy.
After combining these distributions into a likelihood as a function of the fraction of the astrophysical flux that is of Galactic origin $\fgal$, we maximize the likelihood and find $\fgal=0.013$, and then evaluate the probability that each event is background, Galactic, or extragalactic.

The Galactic mass model \cite{McMillan:2011wd} we use for the Galactic contribution to the neutrino flux may not capture the distribution of high energy sources in the Galaxy.
However, due to the large angular uncertainties of the majority of the events, IceCube is unable to discriminate among subtle distinctions.
To verify this, we considered two alternative Galactic distributions -- one of supernova remnants (SNR) \cite{Case:1998qg} and one of pulsar wind nebulae (PWN) \cite{Lorimer:2006qs}.
These different Galactic templates lead to the same $\hfgal=0.013$ for both SNR and PWN.
We conclude that IceCube presently lacks the power to differentiate these models with the HESE data.
As a consequence, our results are independent of the finer details of anisotropy of the Galactic plane.
On the other hand, we have not considered the possible influence of sources outside of our definition of the Galaxy, such as Fermi bubbles, the Crab Nebula, or an extended dark matter halo due to strong limits on these particular structures elsewhere in the literature \cite{Ahlers:2015moa,IceCube:2011aa}.
More data are needed for the evaluation of these further possibilities.

\newcommand{\fluxunits}{GeV cm$^{-2}$ s$^{-1}$ sr$^{-1}$}
The best fit power law to the total astrophysical signal is $E^2\Phi(E)=2.46\pm0.8\e{-8}(E/100$ TeV$)^{-0.92}$ \fluxunits.
For $\hfgal=0.013$, this corresponds to $E^2\Phi_{\rm exgal}(E)=2.43\e{-8}(E/100$ TeV$)^{-0.92}$ \fluxunits~from extragalactic sources and $E^2\Phi_{\rm gal}(E)=3.3\e{-10}(E/100$ TeV$)^{-0.92}$ \fluxunits~from Galactic sources.
If we consider IceCube's best fit broken power law from \cite{IC:ICRC17} $\Phi_{\rm BPL}=\Phi_s+\Phi_h$ with $E^2\Phi_s(E)=1.8\e{-8}(E/100$ TeV$)^{-1.7}$ \fluxunits~and $E^2\Phi_h=8\e{-9}(E/100$ TeV$)^{-0.3}$ \fluxunits.
The extragalactic component is then $E^2\Phi_{{\rm exgal},s}(E)=1.8\e{-8}(E/100$ TeV$)^{-1.7}$ \fluxunits~and $E^2\Phi_{{\rm exgal},h}=7.9\e{-9}(E/100$ TeV$)^{-0.3}$ \fluxunits~while the galactic component is $E^2\Phi_{{\rm gal},s}(E)=2.4\e{-10}(E/100$ TeV$)^{-1.7}$ \fluxunits~and $E^2\Phi_{{\rm gal},h}=1.1\e{-10}(E/100$ TeV$)^{-0.3}$ \fluxunits.

It may be difficult to have enough energetics in our own Galaxy to reach the PeV portion of the observed spectrum \cite{Sahakyan:2015bgg,Dey:2017kly}.
With this in mind, we considered the additional cut $\L_{{\rm gal|astro},i}=0$ for $E_{\nu,i}>900$ TeV.
With this cut in place, events \#14 and \#35 ($p_{\rm gal}=0.36$ and $0.0096$ respectively) become certainly extragalactic. 
The best fit point $\hfgal$ becomes zero, as event \#14 carried most of the likelihood weighting for Galactic events;
we conclude that the evidence for a Galactic contribution to the astrophysical neutrino flux above 60 TeV is very weak.

This result is generally consistent with some other analyses which have found either evidence for a small Galactic contribution \cite{Ahlers:2015moa,Murase:2015xka,Neronov:2015osa,Palladino:2016zoe,Padovani:2016wwn}, or little to no evidence for a Galactic contribution \cite{Ahlers:2013xia,Aartsen:2014gkd,Troitsky:2015cnk,Chianese:2016opp,Palladino:2016xsy,Albert:2017fvi,IC:ICRC17}.
Another analysis focused on energetics alone without anisotropy information found that a large Galactic component is consistent with IceCube's flux measurement \cite{Anchordoqui:2014rca}.
The method presented here benefits from its generality which limits penalty factors for scanning over parameters, or considering numerous possible galactic catalogs, all of which tend to have similar shapes to within the angular resolution of IceCube.
We have quantified here specifically which events are likely to be galactic, and which are likely to be extragalactic.
Corresponding flux comparisons with particular catalogs should be scaled according to whether the catalogs are galactic or extragalactic.

~

\textbf{Note:} Recently IceCube performed an updated analysis of the galactic contribution to the astrophysical neutrino flux \cite{Aartsen:2017ujz}.
Their analysis uses the seven year track data and concludes with a 90\% upper limit on the galactic contribution of $14\%$.

\section{Acknowledgements}
We thank Markus Ahlers, Sergio Palomares-Ruiz, Mohamed Rameez, Irene Tamborra, and Meng-Ru Wu for useful discussions.
PBD acknowledges support from the Villum Foundation (Project No.~13164), and by the Danish National Research Foundation (DNRF91).
DM and TJW acknowledge partial support from the U.S. Department of Energy (DoE) Grant Nos.~DE-SC0010504 and DE-SC0011981, respectively.

\appendix
\section{von Mises-Fisher Distribution}
\label{sec:vMF}
\begin{figure}
\centering
\includegraphics[width=3.5in]{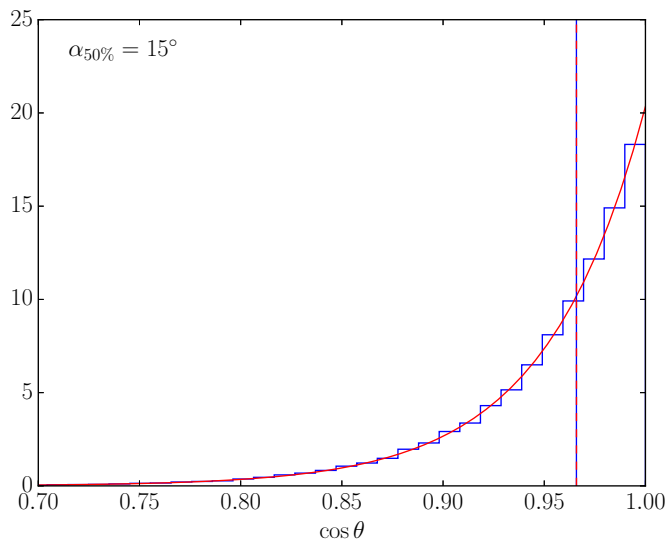}
\caption{The red line is the vMF distribution, eq.~\ref{eq:vMF pdf} for $\kappa=20.3$ scaled by $2\pi$ to include an integral over the azimuthal angle.
The blue histogram is directions sampled according to eq.~\ref{eq:vMF rng} with $\sigma=0.22$.
The blue vertical line is at the median of the sampled points in the blue histogram, while the red dashed vertical line is at $\cos(15^\circ)$.}
\label{fig:vmf}
\end{figure}

A von Mises-Fisher (vMF) distribution is the distribution on a sphere similar to a Gaussian distribution on a line.
For small angular uncertainties ($\lesssim$ few degrees) a Gaussian is sufficient to represent the data, but for larger angular uncertainties (such as the $\sim10^\circ-25^\circ$ of the HESE cascade events), the distinction becomes significant.
A vMF distribution is given simply by the probability density function,
\begin{equation}
f_{\rm vMF}(\theta;\kappa)=\frac{e^{\kappa\cos\theta}}{4\pi\sinh\kappa}\,,
\label{eq:vMF pdf}
\end{equation}
where $\theta$ is the angle away from the center of the distribution and $\kappa$ is the concentration.
$\kappa$ is related to the angular uncertainty $\sigma$ by,
\begin{equation}
1-\kappa\sigma^2=e^{-2\kappa}\,,
\end{equation}
which is well approximated by $\kappa=\frac1{\sigma^2}$ for angular uncertainties $\lesssim30^\circ$.
The angular uncertainty is related to the median spread reported by IceCube by \cite{Ahlers:2015moa},
\begin{equation}
\cos\alpha_{50\%}=1+\sigma^2\log\left[1-0.5\left(1-e^{-\frac2{\sigma^2}}\right)\right]\,.
\end{equation}
This then provides a prescription to generate directions on a sphere distributed according to a vMF distribution,
\begin{equation}
\cos\theta=1+\sigma^2\log\left[1-u\left(1-e^{-\frac2{\sigma^2}}\right)\right]\,,
\label{eq:vMF rng}
\end{equation}
where $\theta$ is the angle away from the central direction, $u\in[0,1)$ is a uniform random variable, and the azimuthal angle around the central direction is uniformly sampled in $[0,2\pi)$.
This is numerically demonstrated in fig.~\ref{fig:vmf}.

\bibliographystyle{JHEP}
\bibliography{IceCube_Anisotropy}

\providecommand{\href}[2]{#2}\begingroup\raggedright\begin{thebibliography}{10}

\bibitem{Aartsen:2013bka}
{\scshape IceCube} collaboration, M.~G. Aartsen et~al., \emph{{First
  observation of PeV-energy neutrinos with IceCube}},
  \href{http://dx.doi.org/10.1103/PhysRevLett.111.021103}{\emph{Phys. Rev.
  Lett.} {\bf 111} (2013) 021103}, [\href{https://arxiv.org/abs/1304.5356}{{\tt
  1304.5356}}].

\bibitem{Anchordoqui:2013dnh}
L.~A. Anchordoqui et~al., \emph{{Cosmic Neutrino Pevatrons: A Brand New Pathway
  to Astronomy, Astrophysics, and Particle Physics}},
  \href{http://dx.doi.org/10.1016/j.jheap.2014.01.001}{\emph{JHEAp} {\bf 1-2}
  (2014) 1--30}, [\href{https://arxiv.org/abs/1312.6587}{{\tt 1312.6587}}].

\bibitem{Aartsen:2014gkd}
{\scshape IceCube} collaboration, M.~G. Aartsen et~al., \emph{{Observation of
  High-Energy Astrophysical Neutrinos in Three Years of IceCube Data}},
  \href{http://dx.doi.org/10.1103/PhysRevLett.113.101101}{\emph{Phys. Rev.
  Lett.} {\bf 113} (2014) 101101}, [\href{https://arxiv.org/abs/1405.5303}{{\tt
  1405.5303}}].

\bibitem{Aartsen:2015zva}
{\scshape IceCube} collaboration, M.~G. Aartsen et~al., \emph{{The IceCube
  Neutrino Observatory - Contributions to ICRC 2015 Part II: Atmospheric and
  Astrophysical Diffuse Neutrino Searches of All Flavors}},  in
  \emph{{Proceedings, 34th International Cosmic Ray Conference (ICRC 2015): The
  Hague, The Netherlands, July 30-August 6, 2015}}, 2015.
\newblock \href{https://arxiv.org/abs/1510.05223}{{\tt 1510.05223}}.

\bibitem{IC:ICRC17}
{\scshape IceCube} collaboration, M.~G. Aartsen et~al., \emph{{Observation of
  Astrophysical Neutrinos in Six Years of IceCube Data}},  in
  \emph{{Proceedings, 35th International Cosmic Ray Conference (ICRC 2017):
  Bexco, Busan, Korea, July 12-20, 2017}}, 2017.

\bibitem{Abbasi:2009cv}
{\scshape IceCube} collaboration, R.~Abbasi et~al., \emph{{Extending the search
  for neutrino point sources with IceCube above the horizon}},
  \href{http://dx.doi.org/10.1103/PhysRevLett.103.221102}{\emph{Phys. Rev.
  Lett.} {\bf 103} (2009) 221102}, [\href{https://arxiv.org/abs/0911.2338}{{\tt
  0911.2338}}].

\bibitem{Adrian-Martinez:2014wzf}
{\scshape ANTARES} collaboration, S.~Adrian-Martinez et~al., \emph{{Searches
  for Point-like and extended neutrino sources close to the Galactic Centre
  using the ANTARES neutrino Telescope}},
  \href{http://dx.doi.org/10.1088/2041-8205/786/1/L5}{\emph{Astrophys. J.} {\bf
  786} (2014) L5}, [\href{https://arxiv.org/abs/1402.6182}{{\tt 1402.6182}}].

\bibitem{Aartsen:2014cva}
{\scshape IceCube} collaboration, M.~G. Aartsen et~al., \emph{{Searches for
  Extended and Point-like Neutrino Sources with Four Years of IceCube Data}},
  \href{http://dx.doi.org/10.1088/0004-637X/796/2/109}{\emph{Astrophys. J.}
  {\bf 796} (2014) 109}, [\href{https://arxiv.org/abs/1406.6757}{{\tt
  1406.6757}}].

\bibitem{Adrian-Martinez:2015ver}
{\scshape IceCube, ANTARES} collaboration, S.~Adrian-Martinez et~al.,
  \emph{{The First Combined Search for Neutrino Point-sources in the Southern
  Hemisphere With the Antares and Icecube Neutrino Telescopes}},
  \href{http://dx.doi.org/10.3847/0004-637X/823/1/65}{\emph{Astrophys. J.} {\bf
  823} (2016) 65}, [\href{https://arxiv.org/abs/1511.02149}{{\tt 1511.02149}}].

\bibitem{Aartsen:2016tpb}
{\scshape IceCube} collaboration, M.~G. Aartsen et~al., \emph{{Lowering
  IceCube's Energy Threshold for Point Source Searches in the Southern Sky}},
  \href{http://dx.doi.org/10.3847/2041-8205/824/2/L28}{\emph{Astrophys. J.}
  {\bf 824} (2016) L28}, [\href{https://arxiv.org/abs/1605.00163}{{\tt
  1605.00163}}].

\bibitem{Mertsch:2016hcd}
P.~Mertsch, M.~Rameez and I.~Tamborra, \emph{{Detection prospects for high
  energy neutrino sources from the anisotropic matter distribution in the local
  universe}},
  \href{http://dx.doi.org/10.1088/1475-7516/2017/03/011}{\emph{JCAP} {\bf 1703}
  (2017) 011}, [\href{https://arxiv.org/abs/1612.07311}{{\tt 1612.07311}}].

\bibitem{Bykov:2015nta}
A.~M. Bykov, D.~C. Ellison, P.~E. Gladilin and S.~M. Osipov, \emph{{Ultrahard
  spectra of PeV neutrinos from supernovae in compact star clusters}},
  \href{http://dx.doi.org/10.1093/mnras/stv1606}{\emph{Mon. Not. Roy. Astron.
  Soc.} {\bf 453} (2015) 113--121},
  [\href{https://arxiv.org/abs/1507.04018}{{\tt 1507.04018}}].

\bibitem{Celli:2016uon}
S.~Celli, A.~Palladino and F.~Vissani, \emph{{Neutrinos and $\gamma$-rays from
  the Galactic Center Region after H.E.S.S. multi-TeV measurements}},
  \href{http://dx.doi.org/10.1140/epjc/s10052-017-4635-x}{\emph{Eur. Phys. J.}
  {\bf C77} (2017) 66}, [\href{https://arxiv.org/abs/1604.08791}{{\tt
  1604.08791}}].

\bibitem{Stecker:1978ah}
F.~W. Stecker, \emph{{Diffuse Fluxes of Cosmic High-Energy Neutrinos}},
  \href{http://dx.doi.org/10.1086/156919}{\emph{Astrophys. J.} {\bf 228} (1979)
  919--927}.

\bibitem{Aartsen:2013jdh}
{\scshape IceCube} collaboration, M.~G. Aartsen et~al., \emph{{Evidence for
  High-Energy Extraterrestrial Neutrinos at the IceCube Detector}},
  \href{http://dx.doi.org/10.1126/science.1242856}{\emph{Science} {\bf 342}
  (2013) 1242856}, [\href{https://arxiv.org/abs/1311.5238}{{\tt 1311.5238}}].

\bibitem{Ahlers:2013xia}
M.~Ahlers and K.~Murase, \emph{{Probing the Galactic Origin of the IceCube
  Excess with Gamma-Rays}},
  \href{http://dx.doi.org/10.1103/PhysRevD.90.023010}{\emph{Phys. Rev.} {\bf
  D90} (2014) 023010}, [\href{https://arxiv.org/abs/1309.4077}{{\tt
  1309.4077}}].

\bibitem{Anchordoqui:2014rca}
L.~A. Anchordoqui, H.~Goldberg, T.~C. Paul, L.~H.~M. da~Silva and B.~J. Vlcek,
  \emph{{Estimating the contribution of Galactic sources to the diffuse
  neutrino flux}},
  \href{http://dx.doi.org/10.1103/PhysRevD.90.123010}{\emph{Phys. Rev.} {\bf
  D90} (2014) 123010}, [\href{https://arxiv.org/abs/1410.0348}{{\tt
  1410.0348}}].

\bibitem{Murase:2015xka}
K.~Murase, D.~Guetta and M.~Ahlers, \emph{{Hidden Cosmic-Ray Accelerators as an
  Origin of TeV-PeV Cosmic Neutrinos}},
  \href{http://dx.doi.org/10.1103/PhysRevLett.116.071101}{\emph{Phys. Rev.
  Lett.} {\bf 116} (2016) 071101},
  [\href{https://arxiv.org/abs/1509.00805}{{\tt 1509.00805}}].

\bibitem{Neronov:2015osa}
A.~Neronov and D.~V. Semikoz, \emph{{Evidence for the Galactic contribution to
  the IceCube astrophysical neutrino flux}},
  \href{http://dx.doi.org/10.1016/j.astropartphys.2015.11.002}{\emph{Astropart.
  Phys.} {\bf 75} (2016) 60--63}, [\href{https://arxiv.org/abs/1509.03522}{{\tt
  1509.03522}}].

\bibitem{Troitsky:2015cnk}
S.~Troitsky, \emph{{Search for Galactic disk and halo components in the arrival
  directions of high-energy astrophysical neutrinos}},
  \href{http://dx.doi.org/10.1134/S0021364015240133}{\emph{JETP Lett.} {\bf
  102} (2015) 785--788}, [\href{https://arxiv.org/abs/1511.01708}{{\tt
  1511.01708}}].

\bibitem{Chianese:2016opp}
M.~Chianese, G.~Miele, S.~Morisi and E.~Vitagliano, \emph{{Low energy IceCube
  data and a possible Dark Matter related excess}},
  \href{http://dx.doi.org/10.1016/j.physletb.2016.03.084}{\emph{Phys. Lett.}
  {\bf B757} (2016) 251--256}, [\href{https://arxiv.org/abs/1601.02934}{{\tt
  1601.02934}}].

\bibitem{Padovani:2016wwn}
P.~Padovani, E.~Resconi, P.~Giommi, B.~Arsioli and Y.~L. Chang, \emph{{Extreme
  blazars as counterparts of IceCube astrophysical neutrinos}},
  \href{http://dx.doi.org/10.1093/mnras/stw228}{\emph{Mon. Not. Roy. Astron.
  Soc.} {\bf 457} (2016) 3582--3592},
  [\href{https://arxiv.org/abs/1601.06550}{{\tt 1601.06550}}].

\bibitem{Palladino:2016zoe}
A.~Palladino and F.~Vissani, \emph{{Extragalactic plus Galactic model for
  IceCube neutrino events}},
  \href{http://dx.doi.org/10.3847/0004-637X/826/2/185}{\emph{Astrophys. J.}
  {\bf 826} (2016) 185}, [\href{https://arxiv.org/abs/1601.06678}{{\tt
  1601.06678}}].

\bibitem{Palladino:2016xsy}
A.~Palladino, M.~Spurio and F.~Vissani, \emph{{On the IceCube spectral
  anomaly}}, \href{http://dx.doi.org/10.1088/1475-7516/2016/12/045}{\emph{JCAP}
  {\bf 1612} (2016) 045}, [\href{https://arxiv.org/abs/1610.07015}{{\tt
  1610.07015}}].

\bibitem{Albert:2017fvi}
A.~Albert et~al., \emph{{Model-independent search for neutrino sources with the
  ANTARES neutrino telescope}},  \href{https://arxiv.org/abs/1703.04351}{{\tt
  1703.04351}}.

\bibitem{Ahlers:2015moa}
M.~Ahlers, Y.~Bai, V.~Barger and R.~Lu, \emph{{Galactic neutrinos in the TeV to
  PeV range}}, \href{http://dx.doi.org/10.1103/PhysRevD.93.013009}{\emph{Phys.
  Rev.} {\bf D93} (2016) 013009}, [\href{https://arxiv.org/abs/1505.03156}{{\tt
  1505.03156}}].

\bibitem{Emig:2015dma}
K.~Emig, C.~Lunardini and R.~Windhorst, \emph{{Do high energy astrophysical
  neutrinos trace star formation?}},
  \href{http://dx.doi.org/10.1088/1475-7516/2015/12/029}{\emph{JCAP} {\bf 1512}
  (2015) 029}, [\href{https://arxiv.org/abs/1507.05711}{{\tt 1507.05711}}].

\bibitem{Kistler:2015oae}
M.~D. Kistler, \emph{{On TeV Gamma Rays and the Search for Galactic
  Neutrinos}},  \href{https://arxiv.org/abs/1511.05199}{{\tt 1511.05199}}.

\bibitem{Fang:2015xhg}
K.~Fang, K.~Kotera, K.~Murase and A.~V. Olinto, \emph{{IceCube Constraints on
  Fast-Spinning Pulsars as High-Energy Neutrino Sources}},
  \href{http://dx.doi.org/10.1088/1475-7516/2016/04/010}{\emph{JCAP} {\bf 1604}
  (2016) 010}, [\href{https://arxiv.org/abs/1511.08518}{{\tt 1511.08518}}].

\bibitem{Sahakyan:2015bgg}
N.~Sahakyan, \emph{{Galactic sources of high energy neutrinos: Expectation from
  gamma-ray data}},
  \href{http://dx.doi.org/10.1051/epjconf/201612105005}{\emph{EPJ Web Conf.}
  {\bf 121} (2016) 05005}, [\href{https://arxiv.org/abs/1512.02333}{{\tt
  1512.02333}}].

\bibitem{Marinelli:2016mjo}
A.~Marinelli, D.~Gaggero, D.~Grasso, A.~Urbano and M.~Valli,
  \emph{{Interpretation of astrophysical neutrinos observed by IceCube
  experiment by setting Galactic and extra-Galactic spectral components}},
  \href{http://dx.doi.org/10.1051/epjconf/201611604009}{\emph{EPJ Web Conf.}
  {\bf 116} (2016) 04009}, [\href{https://arxiv.org/abs/1604.05776}{{\tt
  1604.05776}}].

\bibitem{Arguelles:2017atb}
C.~A. Argüelles, A.~Kheirandish and A.~C. Vincent, \emph{{Imaging Galactic
  Dark Matter with High-Energy Cosmic Neutrinos}},
  \href{https://arxiv.org/abs/1703.00451}{{\tt 1703.00451}}.

\bibitem{Yoast-Hull:2017gaj}
T.~M. Yoast-Hull, J.~S. Gallagher, F.~Halzen, A.~Kheirandish and E.~G. Zweibel,
  \emph{{The Gamma-Ray Puzzle in Cygnus X: Implications for High-Energy
  Neutrinos}},  \href{https://arxiv.org/abs/1703.02590}{{\tt 1703.02590}}.

\bibitem{Stecker:1991vm}
F.~W. Stecker, C.~Done, M.~H. Salamon and P.~Sommers, \emph{{High-energy
  neutrinos from active galactic nuclei}},
  \href{http://dx.doi.org/10.1103/PhysRevLett.66.2697}{\emph{Phys. Rev. Lett.}
  {\bf 66} (1991) 2697--2700}.

\bibitem{Stecker:2005hn}
F.~W. Stecker, \emph{{A note on high energy neutrinos from agn cores}},
  \href{http://dx.doi.org/10.1103/PhysRevD.72.107301}{\emph{Phys. Rev.} {\bf
  D72} (2005) 107301}, [\href{https://arxiv.org/abs/astro-ph/0510537}{{\tt
  astro-ph/0510537}}].

\bibitem{Loeb:2006tw}
A.~Loeb and E.~Waxman, \emph{{The Cumulative background of high energy
  neutrinos from starburst galaxies}},
  \href{http://dx.doi.org/10.1088/1475-7516/2006/05/003}{\emph{JCAP} {\bf 0605}
  (2006) 003}, [\href{https://arxiv.org/abs/astro-ph/0601695}{{\tt
  astro-ph/0601695}}].

\bibitem{Anchordoqui:2014yva}
L.~A. Anchordoqui, T.~C. Paul, L.~H.~M. da~Silva, D.~F. Torres and B.~J. Vlcek,
  \emph{{What IceCube data tell us about neutrino emission from star-forming
  galaxies (so far)}},
  \href{http://dx.doi.org/10.1103/PhysRevD.89.127304}{\emph{Phys. Rev.} {\bf
  D89} (2014) 127304}, [\href{https://arxiv.org/abs/1405.7648}{{\tt
  1405.7648}}].

\bibitem{Zandanel:2014pva}
F.~Zandanel, I.~Tamborra, S.~Gabici and S.~Ando, \emph{{High-energy gamma-ray
  and neutrino backgrounds from clusters of galaxies and radio constraints}},
  \href{http://dx.doi.org/10.1051/0004-6361/201425249}{\emph{Astron.
  Astrophys.} {\bf 578} (2015) A32},
  [\href{https://arxiv.org/abs/1410.8697}{{\tt 1410.8697}}].

\bibitem{Bechtol:2015uqb}
K.~Bechtol, M.~Ahlers, M.~Di~Mauro, M.~Ajello and J.~Vandenbroucke,
  \emph{{Evidence against star-forming galaxies as the dominant source of
  IceCube neutrinos}},
  \href{http://dx.doi.org/10.3847/1538-4357/836/1/47}{\emph{Astrophys. J.} {\bf
  836} (2017) 47}, [\href{https://arxiv.org/abs/1511.00688}{{\tt 1511.00688}}].

\bibitem{Kistler:2015ywn}
M.~D. Kistler, \emph{{Problems and Prospects from a Flood of Extragalactic TeV
  Neutrinos in IceCube}},  \href{https://arxiv.org/abs/1511.01530}{{\tt
  1511.01530}}.

\bibitem{Murase:2015ndr}
K.~Murase, \emph{{Active Galactic Nuclei as High-Energy Neutrino Sources}},
  \href{https://arxiv.org/abs/1511.01590}{{\tt 1511.01590}}.

\bibitem{Aartsen:2016qcr}
{\scshape IceCube} collaboration, M.~G. Aartsen et~al., \emph{{An All-Sky
  Search for Three Flavors of Neutrinos from Gamma-Ray Bursts with the IceCube
  Neutrino Observatory}},
  \href{http://dx.doi.org/10.3847/0004-637X/824/2/115}{\emph{Astrophys. J.}
  {\bf 824} (2016) 115}, [\href{https://arxiv.org/abs/1601.06484}{{\tt
  1601.06484}}].

\bibitem{Moharana:2016yoy}
R.~Moharana, R.~J.~G. Britto and S.~Razzaque, \emph{{Search for extragalactic
  astrophysical counterparts of IceCube neutrino events}}, {\emph{PoS} {\bf
  ICRC2015} (2016) 1122}, [\href{https://arxiv.org/abs/1602.03694}{{\tt
  1602.03694}}].

\bibitem{Murase:2016gly}
K.~Murase and E.~Waxman, \emph{{Constraining High-Energy Cosmic Neutrino
  Sources: Implications and Prospects}},
  \href{http://dx.doi.org/10.1103/PhysRevD.94.103006}{\emph{Phys. Rev.} {\bf
  D94} (2016) 103006}, [\href{https://arxiv.org/abs/1607.01601}{{\tt
  1607.01601}}].

\bibitem{Feyereisen:2016fzb}
M.~R. Feyereisen, I.~Tamborra and S.~Ando, \emph{{One-point fluctuation
  analysis of the high-energy neutrino sky}},
  \href{https://arxiv.org/abs/1610.01607}{{\tt 1610.01607}}.

\bibitem{Neronov:2016ksj}
A.~Neronov, D.~V. Semikoz and K.~Ptitsyna, \emph{{Strong constraint on hadronic
  models of blazar activity from Fermi and IceCube stacking analysis}},
  \href{https://arxiv.org/abs/1611.06338}{{\tt 1611.06338}}.

\bibitem{Aartsen:2017wea}
M.~G. Aartsen et~al., \emph{{Extending the search for muon neutrinos coincident
  with gamma-ray bursts in IceCube data}},
  \href{https://arxiv.org/abs/1702.06868}{{\tt 1702.06868}}.

\bibitem{Aartsen:2015dml}
{\scshape IceCube, Pierre Auger, Telescope Array} collaboration, M.~G. Aartsen
  et~al., \emph{{Search for correlations between the arrival directions of
  IceCube neutrino events and ultrahigh-energy cosmic rays detected by the
  Pierre Auger Observatory and the Telescope Array}}, {\emph{Submitted to:
  JCAP} (2015) }, [\href{https://arxiv.org/abs/1511.09408}{{\tt 1511.09408}}].

\bibitem{Kadler:2016ygj}
M.~Kadler et~al., \emph{{Coincidence of a high-fluence blazar outburst with a
  PeV-energy neutrino event}}, \href{http://dx.doi.org/10.1038/nphys3715,
  10.1038/NPHYS3715}{\emph{Nature Phys.} {\bf 12} (2016) 807--814},
  [\href{https://arxiv.org/abs/1602.02012}{{\tt 1602.02012}}].

\bibitem{Halzen:2016uaj}
F.~Halzen and A.~Kheirandish, \emph{{High Energy Neutrinos from Recent Blazar
  Flares}},
  \href{http://dx.doi.org/10.3847/0004-637X/831/1/12}{\emph{Astrophys. J.} {\bf
  831} (2016) 12}, [\href{https://arxiv.org/abs/1605.06119}{{\tt 1605.06119}}].

\bibitem{Kun:2016bnk}
{Kun, E. and Biermann, P. L. and Gergely, L.Á.}, \emph{{A flat spectrum
  candidate for a track-type high energy neutrino emission event, the case of
  blazar PKS 0723-008}},  \href{https://arxiv.org/abs/1607.04041}{{\tt
  1607.04041}}.

\bibitem{Gao:2016uld}
S.~Gao, M.~Pohl and W.~Winter, \emph{{On the direct correlation between
  gamma-rays and PeV neutrinos from blazars}},
  \href{https://arxiv.org/abs/1610.05306}{{\tt 1610.05306}}.

\bibitem{Abeysekara:2017wzt}
A.~U. Abeysekara et~al., \emph{{Search for Very High Energy Gamma Rays from the
  Northern $\textit{Fermi}$ Bubble Region with HAWC}},
  \href{https://arxiv.org/abs/1703.01344}{{\tt 1703.01344}}.

\bibitem{IceCube:2011aa}
{\scshape IceCube} collaboration, R.~Abbasi et~al., \emph{{Neutrino analysis of
  the September 2010 Crab Nebula flare and time-integrated constraints on
  neutrino emission from the Crab using IceCube}},
  \href{http://dx.doi.org/10.1088/0004-637X/745/1/45}{\emph{Astrophys. J.} {\bf
  745} (2012) 45}, [\href{https://arxiv.org/abs/1106.3484}{{\tt 1106.3484}}].

\bibitem{peter_b_denton_2017_438675}
P.~B. Denton, ``{PeterDenton/ANA v1.0.0}.''
  \href{https://doi.org/10.5281/zenodo.438675}{10.5281/zenodo.438675}, Mar.,
  2017.
\newblock
  \href{https://github.com/PeterDenton/ANA}{github.com/PeterDenton/ANA}.

\bibitem{Glashow:1960zz}
S.~L. Glashow, \emph{{Resonant Scattering of Antineutrinos}},
  \href{http://dx.doi.org/10.1103/PhysRev.118.316}{\emph{Phys. Rev.} {\bf 118}
  (1960) 316--317}.

\bibitem{Anchordoqui:2016ewn}
L.~A. Anchordoqui, M.~M. Block, L.~Durand, P.~Ha, J.~F. Soriano and T.~J.
  Weiler, \emph{{Evidence for a break in the spectrum of astrophysical
  neutrinos}},  \href{https://arxiv.org/abs/1611.07905}{{\tt 1611.07905}}.

\bibitem{Palomares-Ruiz:2015mka}
S.~Palomares-Ruiz, A.~C. Vincent and O.~Mena, \emph{{Spectral analysis of the
  high-energy IceCube neutrinos}},
  \href{http://dx.doi.org/10.1103/PhysRevD.91.103008}{\emph{Phys. Rev.} {\bf
  D91} (2015) 103008}, [\href{https://arxiv.org/abs/1502.02649}{{\tt
  1502.02649}}].

\bibitem{Aartsen:2016xlq}
{\scshape IceCube} collaboration, M.~G. Aartsen et~al., \emph{{Observation and
  Characterization of a Cosmic Muon Neutrino Flux from the Northern Hemisphere
  using six years of IceCube data}},
  \href{http://dx.doi.org/10.3847/0004-637X/833/1/3}{\emph{Astrophys. J.} {\bf
  833} (2016) 3}, [\href{https://arxiv.org/abs/1607.08006}{{\tt 1607.08006}}].

\bibitem{McMillan:2011wd}
P.~J. McMillan, \emph{{Mass models of the Milky Way}},
  \href{http://dx.doi.org/10.1111/j.1365-2966.2011.18564.x}{\emph{Mon. Not.
  Roy. Astron. Soc.} {\bf 414} (2011) 2446--2457},
  [\href{https://arxiv.org/abs/1102.4340}{{\tt 1102.4340}}].

\bibitem{Sergio:2017}
S.~Palomares-Ruiz, ``{Private communication, in which he pointed out to us the
  relevance of this additional factor of $1/4\pi$}.''

\bibitem{wilks1938}
S.~S. Wilks, \emph{The large-sample distribution of the likelihood ratio for
  testing composite hypotheses},
  \href{http://dx.doi.org/10.1214/aoms/1177732360}{\emph{Ann. Math. Statist.}
  {\bf 9} (03, 1938) 60--62}.

\bibitem{Case:1998qg}
G.~L. Case and D.~Bhattacharya, \emph{{A new sigma-d relation and its
  application to the galactic supernova remnant distribution}},
  \href{http://dx.doi.org/10.1086/306089}{\emph{Astrophys. J.} {\bf 504} (1998)
  761}, [\href{https://arxiv.org/abs/astro-ph/9807162}{{\tt
  astro-ph/9807162}}].

\bibitem{Lorimer:2006qs}
D.~R. Lorimer et~al., \emph{{The Parkes multibeam pulsar survey: VI. Discovery
  and timing of 142 pulsars and a Galactic population analysis}},
  \href{http://dx.doi.org/10.1111/j.1365-2966.2006.10887.x}{\emph{Mon. Not.
  Roy. Astron. Soc.} {\bf 372} (2006) 777--800},
  [\href{https://arxiv.org/abs/astro-ph/0607640}{{\tt astro-ph/0607640}}].

\bibitem{Dey:2017kly}
R.~K. Dey, \emph{{PeV neutrinos from local magnetars}},  in \emph{{22nd
  DAE-BRNS High Energy Physics Symposium Delhi, India, December 12-16, 2016}},
  2017.
\newblock \href{https://arxiv.org/abs/1702.01928}{{\tt 1702.01928}}.

\bibitem{Aartsen:2017ujz}
{\scshape IceCube} collaboration, M.~G. Aartsen et~al., \emph{{Constraints on
  Galactic Neutrino Emission with Seven Years of IceCube Data}},
  \href{https://arxiv.org/abs/1707.03416}{{\tt 1707.03416}}.

\end{thebibliography}\endgroup

\end{document}